\documentstyle[prl,aps,preprint]{revtex}

\begin{document}
\title{A Vector Non-abelian Chern-Simons Duality}
\author{H. Garc\'{\i}a-Compe\'an$^{a}$\thanks{
E-mail address: {\tt compean@fis.cinvestav.mx}}, O. Obreg\'on$^{b}$\thanks{
E-mail address: {\tt octavio@ifug3.ugto.mx}} and C. Ram\'{\i}rez$^c$\thanks{
E-mail address: {\tt cramirez@fcfm.buap.mx}}}
\address{$^{a}$ {\it Departamento de F\'{\i}sica\\
Centro de Investigaci\'on y de Estudios Avanzados del IPN}\\
P.O. Box 14-740, 07000, M\'exico D.F., M\'exico\\
$^b$ {\it Instituto de F\'{\i}sica de la Universidad de Guanajuato}\\
P.O. Box E-143, 37150, Le\'on Gto., M\'exico\\
$^c$ {\it Facultad de Ciencias F\'{\i}sico Matem\'aticas\\
Universidad Aut\'onoma de Puebla}\\
P.O. Box 1364, 72000, Puebla, M\'exico}
\date{\today}
\maketitle
\vskip -.5truecm
\begin{abstract}
\vskip -1.4truecm

Abelian Chern-Simons gauge theory is known to possess a  `$S$-self-dual' action where its
coupling constant $k$ is inverted {\it i.e.} $k \leftrightarrow {1 \over k}$. Here a  vector
non-abelian duality is found in the pure non-abelian Chern-Simons action at the classical level. 
The dimensional reduction of the dual Chern-Simons action to two-dimensions constitutes a dual
Wess-Zumino-Witten action already given in the literature.

\end{abstract}

\noindent
hep-th/0103066


\draft



\narrowtext
\newpage

\section{Introduction}

Duality is a very important tool in the study of nonperturbative physics in quantum field and string
theories (for a review, see for instance \cite{giveon}). In this context, duality helps to describe
the strong coupling limit of some supersymmetric field and string theories. Thus, it is important to
determine if a theory does admit dual versions. In order to do that, the Ro\v{c}ek-Verlinde procedure
is very useful \cite{rv}. One general signature to know whether a system can be described through
`dual' variables is the presence of a {\it global} symmetry. This symmetry can be made local to
construct a more general Lagrangian with additional variables (Lagrange multiplier fields) and a
bigger symmetry. From this {\it parent} Lagrangian, the original Lagrangian and its associated dual
Lagrangian can be obtained. This global symmetry can be {\it abelian} or {\it non-abelian} and,
according to it, the above mentioned dualization procedure is called, {\it abelian} or
{\it non-abelian duality}. Abelian duality is nowadays well understood (for a review see for instance
\cite{review,quevedo}).  However non-abelian duality has a more complicated structure. Non-abelian
duality was originally proposed by de la Ossa and Quevedo in \cite{ossa}. Its global structure was
investigated in \cite{alvarez} and further worked out in \cite{rocek,yolanda}. In particular in  Ref. 
\cite{yolanda}, the structure of non-abelian duality of Wess-Zumino-Witten (WZW) models was
studied in detail at the quantum level.  A non-trivial
generalization of the non-abelian $T$-duality is the Poisson-Lie $T$-duality, which was considered by
Klim\v{c}ik and Severa in a series of papers \cite{ks}.

On the other hand, Chern-Simons gauge theory has been used to describe a wide range of phenomena in
three dimensions. This range from condensed matter systems in low dimensions and particularly in the
fractional quantum Hall effect and superconductivity (see for instance \cite{wilczek}), to
$(2+1)$-dimensional gravity \cite{ed}. On the mathematical side, Chern-Simons theory has been very
useful for constructing knots and links invariants \cite{wittencs}.  The study of duality in the {\it
abelian} Chern-Simons action was firstly introduced in Ref.  \cite{rey} and further studied in Refs.
\cite{lee,bala}. In particular, in Ref. \cite{bala}, the effects of T-duality in the fractional
quantum Hall effect were computed.  This duality works by interchanging the level $k$ of the
Chern-Simons theory to ${1 \over k}$. The generalization to the non-abelian Chern-Simons and
supersymmetric Chern-Simons cases has been worked out in Refs. \cite{sabido,sabidosusy}. In these
papers, it was found that at the classical level, the non-abelian dual theories are also non-abelian
Chern-Simons theories with inverted level, just as in the abelian case. Recently, some new non-abelian
dualities in two-dimensional models were discovered by reducing dimensionally certain
three-dimensional non-abelian dual systems \cite{lindstrom}. These results seem to be relevant to
massive IIA supergravity.  Another recent application of non-abelian duality concerns with the dual
descriptions of Belavin-Polyakov instantons \cite{pinzul}.

A very different duality of the Chern-Simons action has also been discussed by Kapustin and Strassler
in the context of the mirror symmetry of the abelian gauge theory in three dimensions \cite{anton}.
In this latter paper it is found that ${\cal N}=3$ Chern-Simons QED and ${\cal N}=4$ QED are in fact
$S$-dual with the mapping $k \leftrightarrow {1\over k}$. In this case the non-abelian vector duality
generalization remain still as an open problem.

It is well known from Ref. \cite{wittencs} that for compact groups, the quantization of
Chern-Simons gauge theory consists on the finite-dimensional Hilbert state constructed from the
conformal blocks of the associated two-dimensional rational conformal field theory (RCFT). Other
features of CFT, like the conformal anomaly and the duality of conformal blocks, can also be
reinterpreted by means of the Chern-Simons three-dimensional theory \cite{csvira}. It is natural to ask
whether the Ro\v{c}ek-Verlinde procedure for the WZW models can be carried over from the Chern-Simons
theory perspective. In this paper we find a positive answer to this question.

We address the problem of non-abelian duality in the Chern-Simons gauge theory in three dimensions,
with compact and simple gauge group. In the process we derive the non-abelian duality of WZW models
found in Ref. \cite{yolanda}, from the Chern-Simons perspective. To be specific, we exploit the fact
that the Chern-Simons action is invariant under global transformations of the connection in the adjoint
representation. This symmetry is gauged out, and the dual action is then obtained. In order to verify
the proposed nonabelian Chern-Simons duality and its possible consequences, we reduce the parent action
to its 2D counterpart.  It turns out that it coincides with the duality found in Ref. \cite{yolanda},
where consequences of it were computed. The reduction to 2D RCFT may help us to understand properly the
structure of nonabelian duality in three dimensions. In \cite{intri}, mirror symmetry of 3D
Chern-Simons theories with a geometric interpretation, realized as a brane configuration of D-branes
and NS-branes, were found. Our results would be relevant to find a relation of \cite{intri} to the
geometrical WZW models realized as brane configurations \cite{mms}.

This paper is organized as follows: In section II we briefly review the necessary tools of non-abelian
duality (we follow Ref.\cite{quevedo}) which will be useful in the subsequent sections. Sections III
and IV are the main contribution of this paper. In section III basically we find the dual non-abelian
Chern-Simons action. In Section IV we reduce the dual action to two dimensions. Finally in Section V we
give our concluding remarks.

\vskip 2truecm
\section{Non-abelian Duality}

In this section we will briefly recall the basics of non-abelian duality. Non-abelian duality was first
proposed in Ref. \cite{ossa}, in the context of the target space duality in string
theory, and further developed in \cite{alvarez,rocek,yolanda,ks}. The starting point is a given
non-linear sigma model described by a Lagrangian $L$ depending of $M$ world-sheet scalar fields
$X^{M}$ and with non-constant target space metric $G_{MN}(X)$. This metric is assumed to possess a
group of non-abelian isometries ${\bf G}$. Let $n$ be the index denoting the isometric directions.
Then scalar fields transform under the global group ${\bf G}$ as $X^m \to g^m_n X^n$ with $g^m_n \in
{\bf G}$. Following the Ro\v{c}ek-Verlinde procedure, one can gauge out a non-abelian subgroup ${\bf
H}$ of ${\bf G}$, with $\partial X^m \to {\cal D}X^m = \partial X^m + A^{\alpha}(T_{\alpha})^m_n
X^n$. The procedure also incorporates to the action a term $\int tr(\Lambda F)$ where $F = \partial
\bar{A} - \bar{\partial} A + [A, \bar{A}]$ and $\Lambda$ is a two-indices Lagrange multiplier field.
The gauge field is a Lie algebra ($Lie(G)$)-valued field in the adjoint representation of ${\bf H}$.

The partition function is given by

\begin{equation}
Z = \int {{\cal D}X \over V_{\bf G}} \int {\cal D} \Lambda {\cal D} A {\cal D} \bar{A}
exp \bigg\{ -i \bigg( S_{gauged}[X,A,\bar{A}] + \int tr(\Lambda F) \bigg) \bigg\}.
\end{equation}

The original action can, as usual, be found by integration of the Lagrange multiplier
$\Lambda$. The dual theory can be obtained by integrating over the gauge fields $A$ and $\bar{A}$. It
yields

\begin{equation}
Z = \int {\cal D}X   {\cal D} \Lambda \delta [{\cal F}] det {\delta {\cal F} \over \delta \omega}
exp \bigg( -i S'[X,\Lambda]\bigg) det( f^{-1}),
\end{equation}
where ${\cal F}$ is the gauge fixing function, $\omega$ represents the parameters of the group
of isometries, $f$ is a matrix-valued coefficient of the quadratic term in the gauge fields and
$S'[X,\Lambda]$ is given by

\begin{equation}
S'[X,\Lambda] = S[X] - {1 \over 4 \pi \alpha '} \int \bar{J}_{\alpha}(f^{-1})^{\alpha \beta}
J_{\beta},\label{corrientes}
\end{equation}
where $J$ and $\bar{J}$ are currents coupled to $\bar{A}$ and $A$ respectively. 

In the next section we will show that the non-abelian Chern-Simons theory possesses {\it exactly}
this non-abelian duality structure. Thus, it will constitute a new example of this sort of duality.

\vskip 2truecm
\section{Non-abelian Chern-Simons Dual Action}

Consider the pure Chern-Simons action

\begin{equation}
L= {k \over 4 \pi} \int_{\cal M} {\rm Tr} \bigg( A dA + {2 \over 3} A^3
\bigg),\label{chernsimons}
\end{equation} 
where $A$ is a connection on the $G$-bundle $E$ over ${\cal M}$, $G$ is a compact and simple Lie
group and ${\cal M}$ is an oriented arbitrary three-manifold with non-empty boundary $\partial {\cal
M} \not= \emptyset$. `Tr' is an invariant quadratic form on the Lie algebra ${\cal G}=Lie(G)$ of $G$.
The wedge product is omitted from this action.

Let
$\{ T_a \},$ $a= 1, \dots , dim(G),$ be a basis of ${\cal G}= Lie(G)$ with $[T^a,T^b]
= f^{ab}_c T^c$ and `Tr' the diagonal quadratic form Tr$(T_aT_b) =
-2\delta_{ab}$.  In local coordinates of ${\cal M}$, the action (4) is thus written as

\begin{equation}
L = {k \over 4 \pi} \int_{\cal M} \varepsilon^{ijk}  \bigg( A^a_i \partial_j 
A_{ka} + {1 \over 3} f_{abc} A^a_i A^b_j A^c_k \bigg).
\end{equation}

The partition function of this theory is given by

\begin{equation}
Z = \int {\cal D} A \,\,exp \bigg({ik \over 4 \pi} \int_{\cal M}  {\rm Tr} \big( AdA +
{2 \over 3} A^3)\bigg).
\end{equation}

We intend to find a ``dual'' action to (5) using the non-abelian generalization of the
Ro\v{c}ek-Verlinde procedure originally proposed in \cite{ossa} and further developed in
\cite{alvarez,yolanda,oganor,oganor1} (for some reviews see, \cite{review,quevedo}.)  We begin by
noting that the action (5) is invariant under the global transformations

\begin{equation}
A_i \to g^{-1} A_i g,
\end{equation}
where $g$ is a constant element of $G$. As in the standard procedure, we gauge a subgroup $H$ of the above symmetry in the action (5), with
algebra $\cal H$, and introduce the $\cal H$-valued gauge
field $B_i$ to get the parent action:

\begin{equation}
L_D =  \int_{\cal M} \varepsilon^{ijk} \bigg[ {k \over 4 \pi}(A_i^a
\widehat{D}_j A_{ka}
+ {1\over 3} f_{abc} A^a_i A^b_j A^c_k) +{1 \over 2 \pi} \chi^a_i F_{jka}(B) \bigg],
\label{accion}
\end{equation}
where the Lagrange multipliers $\chi^a_i$ and the field strength $F^a_{jk}(B) = \partial_jB^a_k -
\partial_k B^a_j + f^a_{bc} B^b_j B^c_k$ are ${\cal H}$-valued forms on ${\cal M}$, and
$\widehat{D}_i= \partial_i + [B_i, \cdot]$ are the corresponding covariant derivatives with respect
to the $B$-fields.  This action is invariant under the symmetry transformations

\begin{equation}
B_i \to h^{-1} B_i h + h^{-1} \partial_i h, \ \ \ A_i  \to  h^{-1} A_i h,  \ \ \ \ \ \chi \to h \chi  h^{-1}, 
\end{equation}
that is,
\begin{equation}
\widehat{D}_iA = h^{-1}\widehat{D}_iA \ h,
\end{equation}
for any element $h$ of $H$.

The partition function for this system is given by

\begin{equation}
Z = \int {\cal D}A{\cal D}\chi {\cal D}B \exp\bigg( i L_D \bigg).
\end{equation}

Integrating with respect to the Lagrange multipliers $\chi^a_i$, we get the 
constraints

\begin{equation}
F^a_{jk}(B) = 0,
\end{equation}
which lead  to consider only flat gauge connections $B_i$ of the form

\begin{equation}
B_i = h^{-1} \partial_i h.
\end{equation}
Then locally the gauge fields are pure gauge, the gauge fixing $B^a_i = 0$ can be chosen, and we
recover the original action (5).

On the other hand, the ``dual'' action can be obtained by integrating over the ${\cal H}$-valued gauge fields $B^a_i$ and
then fixing the gauge. The relevant part of the action is

\begin{equation}
L_D = \int_{\cal M} \varepsilon^{ijk}\bigg( \dots + {k \over 8 \pi}f_{abc}A^a_i A^b_j B^c_k +{1 \over 2 \pi}
\chi^a_i \partial_j B_{ka} + {1 \over 4 \pi} f_{abc} \chi^a_iB^b_j B^c_k + \dots
\bigg).
\end{equation}
This is a Gaussian integral and the integration defines the ``dual'' action $L^*_D$ with
partition function

\begin{equation}
Z= \int {\cal D} A {\cal D} \chi \det M^{-1/2} exp \bigg( i L^*_D \bigg),\label{dual1}
\end{equation}
with the H-invariant dual action,
\begin{eqnarray}
L^*_D[A,\chi] &=& L[A] + {1 \over 4 \pi}\int_{\cal M} J_i^a(\chi,A) M^{-1\ ij}_{\ \ ab}
J_j^b(\chi,A)\nonumber \\ &+&{1 \over 4 \pi}\int_{\partial {\cal M}} N_a^i(\chi) M^{-1\ ab}_{\ \ ij} \bigg(N^j_b(\chi)-2 J^j_b(\chi,A) \bigg),
\label{resultadocs}
\end{eqnarray}
where $L[A]$ is the standard Chern-Simons action (5), 
\begin{equation}
J^k_a(\chi,A) = \varepsilon^{ijk} \bigg( \partial_i \chi_{ja} + {k \over 8 \pi} f_{abc}
A^b_i A^c_j \bigg),
\end{equation}
$N^i_a = \varepsilon^{ijk} n_j \chi_{ka}$, with $n_i$ the normal to $\partial {\cal M}$, 
and $ M^{-1}$ is the inverse matrix of \begin{equation}
M^{ij}_{ab} = {1 \over 2} \varepsilon^{ijk} f_{abc} \chi^c_k.
\label{eme}
\end{equation} 
Up to boundary terms, the form of the dual action (\ref{resultadocs}) coincides with
(\ref{corrientes}), with the difference that the factor determinant in (\ref{dual1}), appears here 
with a square root due to the fact that the integral is Gaussian.  The action (\ref{resultadocs}) is
still gauge invariant under $H$ transformations and as usual, in order to obtain the true dual
action, a gauge fixing has to be undertaken. However, as pointed out in \cite{ossa} the dual action
shall not depend the definite form of the gauge fixing.

The matrix (\ref{eme}) can be in general singular. Algebraically its inverse can be given, although
not in a general simple form. For example for $H$=SU(2) it is given by 
$$M^{-1\ ab}_{\ \ ij}={1\over
det(M)}(\chi_i^a \chi_j^b-2\chi_j^a \chi_i^b).$$ 
In the case $det(M)$ has singularities, counterterms
have to be added in order to regularize the corresponding poles \cite{oganor}.

\vskip 2truecm
\section{Dual Wess-Zumino-Witten Action}

In order to understand the precise structure of the three-dimensional dual action (16), we compute
the corresponding theory on the boundary $\partial {\cal M}$ of ${\cal M}$. To do this we
follow Refs. \cite{zoo,elitzur}. If the manifold ${\cal M}$ has a boundary, we can consider the two
dimensional theory corresponding to the action (\ref{accion}). In order to do that, we separate the
time from the space components: $d=d_0+\widetilde{d}$, $ A=A_0+\widetilde{A}$, $B=B_0+\widetilde{B}$
and $\chi = \chi_0 + \widetilde{\chi}$, then the parent action (8) can be rewritten as

\begin{eqnarray} 
L_D= \int_{\cal M}{1\over 4\pi}
&{\rm Tr} &\bigg[-{1\over{2}}\widetilde{A}d_0\widetilde{A}+ \widetilde{\chi}d_0\widetilde{A}+\chi_0
\widetilde{F}(\widetilde{B})-
k A_0 (\widetilde{d}\widetilde{A}+ \widetilde{B}\widetilde{A}+ \widetilde{A}\widetilde{B}+\widetilde{A}^2)+
\nonumber
\\
&&B_0(\widetilde{d}\widetilde{\chi}+\widetilde{B}\widetilde{\chi}+\widetilde{\chi}\widetilde{B}-k
\widetilde{A}^2)\bigg], 
\end{eqnarray} 
where we set the boundary conditions $A_0=B_0=0$.

Thus, after integrating out the Lagrange multipliers, we have the following equations:
\begin{equation}
\widetilde{F}(\widetilde{B})=\widetilde{d}\widetilde{B}+\widetilde{B}^2=0,
\end{equation}
\begin{equation}
G(\widetilde{A},\widetilde{B})\equiv
\widetilde{d}\widetilde{A}+\widetilde{B}\widetilde{A}+\widetilde{A}\widetilde{B}+\widetilde{A}^2=0,
\end{equation}
\begin{equation}
H(\widetilde{A},\widetilde{B},\widetilde{\chi})\equiv
\widetilde{d}\widetilde{\chi}+\widetilde{B}\widetilde{\chi}+\widetilde{\chi}\widetilde{B}-k\widetilde{A}^2=0.
\end{equation}

Further, we observe that if we define $\widetilde{A}= \widetilde{\phi}- \widetilde{B}$ and
$\widetilde{\chi}=-k \widetilde{A}+\widetilde{\lambda}$, where $\widetilde{\phi}\in{\cal G}$ and $\widetilde{\lambda}\in{\cal H}$, then
$G(\widetilde{A},\widetilde{B})=\widetilde{F}(\widetilde{\phi})-\widetilde{F}(\widetilde{B})$ and $
H(\widetilde{A},\widetilde{B},\widetilde{\chi})=-k G(\widetilde{A},\widetilde{B})
+\widetilde{d}\widetilde{\lambda}+
\widetilde{B}\widetilde{\lambda}+\widetilde{\lambda}\widetilde{B}$. Therefore the equations we have to
set to zero are
$\widetilde{F}(\widetilde{B})=F(\widetilde{\phi})=0$ and $\widetilde{d}\widetilde{\lambda}+
\widetilde{B}\widetilde{\lambda}+\widetilde{\lambda}\widetilde{B}=0$. The two first equations can be
solved by
$\widetilde{B}=h^{-1}\widetilde{d}h$ and $\widetilde{\phi}=g^{-1}\widetilde{d}g$, where $h\in H$ and $g\in G$. Now, if we insert these solutions into the last equation, we get,

\begin{equation}
\widetilde{d} \widetilde{\lambda}+ h^{-1}\widetilde{d}h \widetilde{\lambda}+ \widetilde{\lambda}
h^{-1}\widetilde{d}h=0
\end{equation}
that is
\begin{equation}
h\widetilde{d}\widetilde{\lambda} h^{-1}+ \widetilde{d}h \widetilde{\lambda} h^{-1}+h
\widetilde{\lambda}
h^{-1}\widetilde{d}h
h^{-1}=\widetilde{d}(h\widetilde{\lambda} h^{-1})=0.
\end{equation}
Therefore 
\begin{equation}
\widetilde{\lambda}= h^{-1}\widetilde{d}\alpha \ h, \ \ \widetilde{\chi}=-k\widetilde{A}+
h^{-1}\widetilde{d}\alpha \ h
\label{alfa}
\end{equation}
Where $\alpha\in {\cal H}$. We get,
\begin{equation}
L_D=\int_{\cal M}{1\over 4\pi}{\rm Tr}\bigg[{k\over{2}}g^{-1}\widetilde{d}g
d_0(g^{-1}\widetilde{d}g) -{k\over{2}} h^{-1}\widetilde{d}h
d_0(h^{-1}\widetilde{d}h)+ h^{-1}\widetilde{d}\alpha h d_0(h^{-1}\widetilde{d}h)\bigg],
\end{equation}
which can be rewritten as
\begin{equation}
L_D=\int_{\cal M} {1\over 4\pi}{\rm Tr}\bigg[\widetilde{d}\bigg({k\over{2}}\widetilde{d}g^{-1}d_0
g-{k\over{2}}\widetilde{d}h^{-1}d_0
h+\widetilde{d}\alpha d_0 h h^{-1}\bigg)-{k\over{6}} (h^{-1}dh)^3+{k\over{6}} (g^{-1}dg)^3 \bigg].
\end{equation}

Therefore, if for example our manifold is ${\cal M}={\bf R}\times {\bf D}$, where ${\bf D}$ is a $2-$disk,
then
if $r$ and $\phi$
are the coordinates on the disk, we get the two dimensional parent action,

\begin{eqnarray}
I_D=&& \int_{{\bf R} \times \partial {\bf D}} {1\over 4\pi} {\rm Tr}
\bigg(-{k\over{2}}g^{-1}\partial_{\phi}g
g^{-1} \partial_{t}g +
{k\over{2}}h^{-1}\partial_{\phi}h h^{-1} \partial_{t}h -\partial_{\phi}\alpha \partial_t h
h^{-1}\bigg) d\phi dt \nonumber \\ &&+ {k\over{24\pi}} \int_{{\bf R} \times {\bf D}}
{\rm Tr}\bigg[(g^{-1}dg)^3-
(h^{-1}dh)^3 \bigg].
\label{parienteuno}
\end{eqnarray}

This parent action contains two WZW actions for $g$ and $h$, as well as the $\alpha$-term. It coincides
with the nonabelian duality parent action for WZW given in \cite{yolanda} (see Eq. (4.16) of Ref.
\cite{yolanda}).

From (\ref{alfa}), we see that the field $\alpha$ corresponds to the field $\chi$ in (\ref{accion}) and
thus its integration should give the WZW action as result. Indeed, the integration over $\alpha$ gives
$\partial_\phi(h^{-1}\partial_t h)=0$, whose solution is $h(t,\phi)=A(\phi)B(t)$. After substitution of
this solution back into the parent action (\ref{parienteuno}), the $\alpha$-term and the $h$ WZW action
vanish identically. Thus, as expected, the resulting action is the WZW action corresponding to the
Chern-Simons action (\ref{chernsimons}),

\begin{equation}
I_{WZW} ={k\over 2}\int {\rm Tr} \ g^{-1}\partial_{\phi}g g^{-1}\partial_{t}g d\phi dt+{k\over 6}
{\rm Tr}\int
(g^{-1}dg)^3.
\label{accionprimaria}
\end{equation}

In order to get the dual action to (\ref{accionprimaria}), it has to be integrated over the $h$ field
in (\ref{parienteuno}) \cite{yolanda}. Let us denote by $I_h$ the action containing $h$, its variation
with respect to $h$ gives

\begin{equation}
\delta_h I_h=-{1\over 4\pi} \int {\rm Tr} \bigg[h^{-1}\delta h \bigg( \partial_t\partial_\phi\alpha
-k\partial_\phi(h^{-1}\partial_t
h)+ [ h^{-1} \partial_t h,\partial_\phi\alpha]\bigg)\bigg] dt d\phi=0,
\label{accion1}
\end{equation}
which is solved by 
\begin{equation}
\partial_\phi\alpha=k h^{-1} \partial_\phi h,\label{alfa1}
\end{equation}
whose substitution in the parent action gives,
\begin{eqnarray}
\widetilde{I}_{WZW}=&&-  {1\over 4\pi}\int_{{\bf R} \times \partial {\bf D}} {\rm Tr}
\bigg({k\over{2}}g^{-1}\partial_{\phi}g
g^{-1} \partial_{t}g +
{k\over{2}}h^{-1}\partial_{\phi}h h^{-1} \partial_{t}h\bigg) d\phi dt \nonumber \\ &&+ {k\over{24\pi}} \int_{{\bf R} \times {\bf D}}
{\rm Tr}\bigg[(g^{-1}dg)^3-
(h^{-1}dh)^3 \bigg].
\label{parientedos}
\end{eqnarray}

A detailed analysis done in \cite{yolanda}, taking into account quantum corrections, in particular the
ones due to the change of variable (\ref{alfa1}), shows that this dual nonabelian WZW action
corresponds exactly to a $(G/H)_k \times H_k$ WZW model.

\vskip 2truecm
\section{Concluding Remarks}

In this paper we have further investigated the structure of non-abelian duality. We have found that one
can associate a dual action to the non-abelian Chern-Simons action, which constitutes a new example of
this kind of duality. After solving the constraints, we find an explicit `dual' action for the original
Chern-Simons action (5) in terms of a the Lagrange multiplier variables $\chi_{k}^a$, given by Eq.
(16). This action contains a Chern-Simons term of the original fields $A_i^a$ plus the $\chi$ action
and a boundary term. It needs still to be gauge fixed in order to eliminate redundant degrees of
freedom. In order to see which 2D CFT theory corresponds to this dual, we resort to the two
dimensional theory on the boundary ${\bf R} \times \partial {\bf D}$ corresponding to the parent
action. This procedure was performed in the Sec. IV and the non-abelian action $I_D$ obtained in Ref.
\cite{yolanda} turned out.

We can summarize our results with the following diagram:

$$ 
\matrix{
{\rm CS} & {\buildrel{D} \over {\longrightarrow}} & \widetilde{\rm CS}\cr
\downarrow{R}{} && \downarrow{R}&{}&\cr
{\rm WZW} & {\buildrel{D} \over {\longrightarrow}} &  \widetilde{\rm WZW}\cr
}
$$
where the mapping consisting in obtaining the dual action is denoted by $D$. $R$ denotes the dimensional reduction of the dual action on the boundary ${\bf R} \times \partial{\bf D}$. The reduced parent action $I_D$ Eq. (28) corresponds to two coupled WZW actions,
just as was found in Ref.  \cite{yolanda}. It is interesting to see that the above diagram commutes.
The reason of this is that the dimensional reduction of $\widetilde{\rm CS}$ theory, {\it i.e.}
$\widetilde{\rm WZW}$ model, coincides with the dual action $\widetilde{I}_{WZW}$ to the WZW model
obtained in \cite{yolanda} which comes directly from dimensional reduction of the $CS$ theory
according to Refs. \cite{zoo,elitzur} and it is given by Eq. (32). This is so because both approaches
have the same WZW parent action $I_D$ Eq. (28).

It was the main aim of the paper to construct the non-abelian dual theory of the non-abelian
Chern-Simons theory.  Similarly to another examples, the dual action is a Chern-Simons action coupled
to a Freedman-Townsend-like action Eq. (16). To find the utility of the non-abelian Chern-Simons
duality, it remains to apply it to some systems involving non-abelian Chern-Simons theory. One of these
examples would be to find a relation between some properties in the strong/weak coupling region of the
Chern-Simons gauge theory to the weak/strong one of the same theory. This could be of importance, for
instance, relate the topological invariants of knots and links defined in the strong coupling limit ${1
\over k} \to \infty$ (Jones polynomial) and that defined in the weak coupling limit ${1 \over k} \to 0$
(Vassiliev invariants).

Finally, non-abelian Chern-Simons action with non-compact complex groups are relevant in the
description of $(2+1)$ quantum gravity \cite{ed}. Dual actions for gravity and supergravity were found
in \cite{sabido,sabidosusy}. We would like to apply the issues considered here, to the
Chern-Simons (super)gravity case and compare with the dual actions obtained in
\cite{sabido,sabidosusy}. From Ref. \cite{csvira} it is known that for the gravitational case,
the associated Hilbert space is infinite dimensional. Even in this case CFT is of extreme importance 
to describe the gravity system \cite{nat}. Thus non-abelian Chern-Simons duality and its 
two-dimensional reduction would be useful to address the gravitational case. 

Also, it is well known that non-abelian Chern-Simons gauge theory can be
regarded as a topological string theory \cite{csstring}. It is tantalizing to apply the dual
Chern-Simons action in order to look for $S$-duality structure in the various involved topological
sigma models. It would be also interesting to compare our results with that obtained by Mohammedi
\cite{mohammedi}. Some of these subjects are now under current investigation.

\vskip 2truecm 
\centerline{\bf Acknowledgments}

This work was supported in part by CONACyT grants 28454E and 33951E.


\vskip 2truecm 



\end{document}